\documentclass[12pt]{article}
\usepackage[hmargin=2cm,vmargin=1.5cm]{geometry}
\usepackage{graphicx}
\usepackage{amsmath}
\usepackage{enumerate}

\begin{document}
\title{Quantum temporal probabilities in tunneling systems: II. No faster-than-light signals are possible in tunneling.}
\author {Charis Anastopoulos
\footnote{anastop@physics.upatras.gr} and Ntina Savvidou
\footnote{ksavvidou@physics.upatras.gr}
\\ Department of Physics, University of Patras, 26500 Greece}

 \maketitle

\begin{abstract}
  In this article, we propose a resolution to the paradox of apparent superluminal velocities for tunneling particles,
by a careful treatment of temporal observables in quantum theory and through a precise application of the duality between particles and waves. To this end, we employ a new method for constructing probabilities associated to quantum time measurements that provides an explicit link between the tunneling time of particles and the associated quantum fields. We demonstrate that the idea of faster-than-light speeds in tunneling follows from an inadmissible use of classical reasoning in the description of quantum systems. Our results suggest that direct measurements of the transit time in tunneling could provide a new testing ground for the predictions of quantum theory versus local hidden-variables theories.
\end{abstract}

\section{Introduction}

One of the most fundamental features of quantum theory is the duality between particles and waves, namely, the fact that the same quantum system can exhibit particle-like or wave-like properties in different experimental configurations. The particle and the wave aspects of any system are complementary, they cannot be combined into a single unified description of microscopic properties that is independent of the experimental set-up. In this article, we show that the long-standing paradox of apparent super-luminal velocities in quantum tunneling can be resolved by a careful implementation of particle-wave duality in the context of Quantum Field Theory (QFT).

The superluminality paradox in tunneling originates from the early days of quantum theory \cite{old}, in the effort to identify how long it takes a quantum particle to tunnel through a potential barrier. The search for an answer to this question has led to a large number of candidates for the tunneling time \cite{rev, Win06} rather than to a single expression derived unambiguously from first principles. Many existing definitions imply that tunneling times saturate in the opaque-barrier limit, thus suggesting superluminal speeds for particles traversing sufficiently long barriers (the Hartmann effect \cite{Har}).

Several experimenters have reported superluminal tunneling velocities in electromagnetic analogues of quantum tunneling \cite{EN, SKC, SSSK, BD, MSHM, LMLB}. The analogy is based on the mathematical correspondence between the classical Helmholtz equation for the electromagnetic field in inhomogeneous dielectric media and the time-independent Schr\"odinger equation in presence of a potential. The experiments above measure the group delay $t_d$ associated to electromagnetic pulses crossing a barrier of length $d$. In classical field theory, the group delay is standardly interpreted as the transit time of a signal through a medium or a device. Hence, the group velocity $d/t_d$ (which is found greater than the speed of light) is interpreted as the transit velocity for waves crossing the barrier. However, no direct measurement of the transit time is involved in these experiments; the transit time is inferred from the group delay.

Our approach to the superluminality paradox is based on the observation that the relation between group delay and transit time is a feature of classical physics and not of quantum physics. In quantum theory, there is no such relation, because group delay and transit time are incompatible observables that are determined in distinct types of experiment. The former is an observable in experiments that measure the wave aspects of a quantum system, while the latter refers to the propagation of localized excitations of quantum fields, i.e., particles. The interpretation of group delay as a transit time violates a fundamental rule of quantum mechanics, Bohr's complementarity principle, according to which "evidence obtained under different experimental conditions cannot be comprehended within a single picture" \cite{Bohr}.

Group delay in electromagnetic analogues of tunneling has a natural interpretation as the lifetime of energy stored in the barrier \cite{Win06, Win}. To address the issue of superluminality, we must consider experiments in which the particles' transit times (or times of arrival) are direct physical observables.  Such experiments are well understood in classical theory. They typically involve a particle source and a particle detector separated by distance $L$ at rest with respect to each other. Source and detector are equipped with a pair of synchronized clocks. The transit time is defined as the difference $t$ between the clock readings of detection and emission respectively, and the traversal velocity is defined as $L/t$. When considering quantum particles, the time difference $t$ becomes a random variable that may take different values in different runs of the experiment. Thus, a quantum description requires the construction of a probability density $P(L, t)$ for the transit time. When a potential barrier is placed along the line connecting emitter and detector, all information about the temporal aspects of tunneling is contained in the probability density $P(L, t)$ \cite{AnSav08}.

However, the construction of a probability density $P(L, t)$ for transit times associated to tunneling experiments is a highly non-trivial task. The derivation of probabilities with respect to time in quantum theory has run into the same type of problems as the identification of tunneling time \cite{toa}. These problems originate from the subtle role of time in quantum theory, where the time $t$ appearing in Schr\"odinger's equation is an external parameter and not an ordinary observable, like position or momentum. This implies that the squared modulus of the time-evolved wave-function $|\psi(x, t)|^2$ is not a density with respect to $t$, and, hence, it cannot serve as a definition for the required probabilities.

A significant advance towards the resolution of the issues above is the development of an algorithmic method for constructing Quantum Temporal Probabilities (QTP) associated to any experimental configuration \cite{AnSav12}---see also Ref. \cite{many}. The QTP method incorporates the detector degrees of freedom into the quantum description, so that the temporal probabilities are always defined with respect to specific experimental set-ups. The key idea is to distinguish between the roles of time as a parameter to Schr\"odinger's equation and as a label of the causal ordering of events \cite{Sav}. This important distinction allows for the definition of quantum temporal observables. In particular, we identify the time of a detection event as a coarse-grained quasi-classical variable \cite{GeHa}  associated to macroscopic records of observation. Thus, the time variables in QTP correspond to macroscopic observable magnitudes, such as the coincidence of a detector `click' with the reading of a clock external to the system.

 In Ref. \cite{AnSav12b}, we used the QTP method in order to derive of a probability density $P(L, t)$ for the time $t$ of particle detection in tunneling set-ups. The derivation involves a QFT description of the tunneling process that is necessary for a definitive resolution of the superluminality paradox. The reason is that superluminal speeds imply a violation of the principle of local causality, and this principle is implemented in quantum theory only if the interactions are expressed in terms of local quantum fields \cite{qft}.

 In what follows, we briefly present the relevant results of Ref. \cite{AnSav12b}) (Sec.2), and then we use these results in order to propose a resolution of the superluminality paradox (Sec. 3).

\section{The QTP method}

 In Ref. \cite{AnSav12b}, we constructed the detection probability for quantum particles of mass $m$ and charge $e$ , described in terms of a quantum field $\hat{\phi}(x)$ tunneling through a potential barrier. Considering for simplicity a single spatial dimension, we assume that the barrier is localized in a region $D = [-d/2, d/2]$, where $d$ is the length of the barrier. A particle detector is localized around $x = L >> d$. In the QTP method, the explicit quantum modeling of the particle detector leads to a unique probability density $P(t)$ associated to an experiment. The most important element in the modeling of the detector is the specification of the interaction Hamiltonian $\hat{H}_I$  between the detector and the quantum field $\hat{\phi}(x)$. The requirements of Lorentz covariance and unitarity imply that $\hat{H}_I$  must be a local functional of the fields $\hat{\phi}(x)$ and $\hat{\phi}^{\dagger}(x)$ \cite{qft}. We select
\begin{eqnarray}
\hat{H}_{I} = \int dx \left[\hat{\phi}(x) \otimes \hat{J}^{\dagger}(x) + \hat{\phi}^{\dagger}(x) \otimes \hat{J}(x) \right], \label{hint}
\end{eqnarray}
where $\hat{J}$ and $\hat{J}^{\dagger}$ are current operators on the Hilbert ${\cal H}_{det}$ of the detector degrees of freedom. The initial state of the detector $|\Psi_0\rangle $ is assumed to satisfy the condition $\hat{J}(x) |\Psi_0\rangle = 0$, which guarantees that the detector is sensitive to particles rather than anti-particles.

For the set-up described above, we derived a general expression for the probability density $P(L, t)$
 \begin{eqnarray}
P(L, t) = \int d\tau \int dz R(z, \tau)   \langle \psi_0|\hat{\phi}^{\dagger}(L -\frac{z}{2}, t-\frac{\tau}{2}) \hat{\phi}(L+\frac{z}{2}, t+\frac{\tau}{2})|\psi_0\rangle  . \label{prob1}
\end{eqnarray}
where $| \psi_0 \rangle$ is the initial state of the field and $\hat{H}_0$ the corresponding Hamiltonian operator. All information about the detector is contained in a kernel  $R(z, t)$ that smears the field correlation function.  It is important to emphasize that Eq. (\ref{prob1}) requires no specification of the Hamiltonian   $\hat{H}_0$ or the quantum field $\hat{\phi}$. In particular, it does not depend on how we choose to model the potential barrier. It is valid for  a barrier modeled by a background classical field, but also for barriers defined in terms of the full quantum interaction of the field $\hat{\phi}$ with other fields (i.e., the quantum electromagnetic field). The only assumption in the derivation of Eq. (\ref{prob1}) is that the detector at $x= L$  detects particles corresponding to the field $\hat{\phi}$.
Thus,  Eq. (\ref{prob1}) reveals a relation between  probabilities for detection time  and field correlation functions that persists in any quantum field theory.

For concreteness, we specialized to the case that the potential barrier corresponds to a background static electromagnetic potential $A_{\mu}(x) = (A_0(x), 0)$, where $A_0(x)$ differs from zero only in the spatial region $D$. Then, the classical Hamiltonian for the field $\phi(x)$  is
\begin{eqnarray}
H  = \int dx \left[|\pi|^2 + |\partial_x \phi|^2 + m^2 |\phi|^2 + i V (\pi \phi - \pi^* \phi^*) \right], \label{hamd}
\end{eqnarray}
where $\pi(x)$ is the conjugate momentum to $\phi(x)$ and $V(x) = e A_0(x)$. The quantized Hamiltonian $\hat{H}$ is expressed in terms of the standard creation and annihilation operators: $\hat{a}(x), \hat{a}^{\dagger}(x)$ for particles and  $\hat{b}(x), \hat{b}^{\dagger}(x)$ for anti-particles

\begin{eqnarray}
\hat{H} = \int dx dx' \left(\hat{a}^{\dagger}(x)h_1(x,x') \hat{a}(x) + \hat{b}^{\dagger}(x)h_2(x,x') \hat{b}(x)\right), \label{hamilton}
\end{eqnarray}
where $h_1(x, x')$ and $h_2(x, x')$ denote matrix elements of  Hamiltonian operators $h_1$ and $h_2$ on the Hilbert spaces $H_1$ and $H_1^*$ associated to single particle and a single anti-particle, respectively. The operators $h_1$ and $h_2$ are defined as $h_{1,2} = h_0 \pm \tilde{V}(x)$, where $h_0 = \sqrt{-\partial_x^2 + m^2}$  is the Hamiltonian for a single free particle, and  $\tilde{V} := \frac{1}{2} \left( h_0^{1/2} V h_0^{-1/2} + h_0^{1/2} V h_0^{-1/2}\right) = V + [[V,h_0^{1/2}], h_0^{-1/2}]$ is the QFT-corrected potential term.

For a positive-valued potential $V(x)$, the single-particle Hamiltonian $h_1$ has continuous spectrum. There exists a pair of orthogonal energy eigenstates to each value of energy $E>m$. We denote such eigenstates as  $f_{k+}$ and $f_{k-}$, where $k=\sqrt{E^2-m^2} > 0$. The eigenstates $f_{k+}$ correspond to only positive momentum flux at $x = \infty$. They are of the form

\begin{eqnarray}
f_{k+}(x) = \frac{1}{\sqrt{2\pi}}  \left\{ \begin{array}{cc} e^{ikx} + R_k e^{-ikx} & x < -\frac{d}{2} \\
T_k e^{ikx} & x > \frac{d}{2}  \end{array} \right. \label{f+}
\end{eqnarray}

where $T_k$ and $R_k$ are the usual transmission and reflection coefficients for a right-moving plane wave of momentum $k$. We parameterize the eigenstates $f_{k-}(x)$ as
\begin{eqnarray}
f_{k-}(x) = \frac{1}{\sqrt{1 - w_k^2}} \left[ - w_k f_{k+}(x) + f_{k+}(-x) \right],
 \end{eqnarray}
 where $w_k= \frac{1}{2} (T_k R_k^*+T_k^* R_k )$ is essentially the overlap between $f_{k+}$ and its parity transform. For parity-symmetric potentials, $V(x) = V(-x)$, the coefficient $w_k$ vanishes. The Hamiltonian $h_2$ for a single antiparticle, has discrete-spectrum eigenstates $g_n(x)$ for energies $E_n <m$, in addition to the continuous spectrum eigenstates $g_{k±}$ (analogous to the $f_{k±}$ above) for $E>m$.

Defining $\hat{a}_{k\pm} = \int dx \hat{a}(x) f^*_{k\pm}(x)$, $\hat{b}_{k \pm} =  \int dx \hat{b}(x) g^*_{k\pm}(x)$, and $\hat{b}_n = \int dx \hat{b}(x) g^*_{n}(x)$, we express the Heisenberg-picture field operators as $\hat{\phi}(x,t)=  \hat{\phi}_1 (x,t) + \hat{\phi}_2 (x,t)$, where
\begin{eqnarray}
\hat{\phi}_1(x, t) &=& \sum_{\sigma = -}^+ \int_0^{\infty} \frac{dk}{\sqrt{2E_k}} \hat{a}_{k,\sigma} f_{k, \sigma}(x) e^{-i E_k t} \label{fi1}
 \\
 \hat{\phi}_2(x,t) &=& \sum_{\sigma = -}^+ \int_0^{\infty} \frac{dk}{\sqrt{2E_k}}
 \hat{b}^{\dagger}_{k,\sigma} g^*_{k, \sigma}(x) e^{i E_kt} \nonumber \\ &+& \sum_n \frac{1}{\sqrt{2E_n}} \hat{b}^{\dagger}_n g_n^*(x) e^{iE_nt} \label{fi2}
\end{eqnarray}

We assume that the particles are emitted from a source, localized around $x = -x_0 < -d/2$, in an initial state $|\psi_0 \rangle = \int dx \hat{a}^{\dagger}(x) \psi_0(x) |0\rangle$, where $\psi_0(x)$ is a single-particle wave-function with positive momentum and $|0 \rangle$ is the field vacuum. Substituting Eqs. (\ref{fi1}---\ref{fi2}) into Eq. (\ref{prob1}), we find that the contribution from the anti-particle component $\hat{\phi}_2$ is negligible for $L >> d$. For the detector far from the barrier, the probability density $P(L< t)$ becomes \cite{AnSav12}

 \begin{eqnarray}
P(L, t) = \left| \int_0^{\infty} \frac{dk}{2\pi} \sqrt{\alpha(k) |v_k|} A_k \tilde{\psi}_0(k) e^{ikL-i E_k t}\right|^2. \label{plt}
\end{eqnarray}
where $v_k=k/\sqrt{k^2+m^2}$ is the relativistic velocity and $\psi_0 (k)$ is the initial wave-function in the momentum representation. In Eq. (\ref{plt}), all information about the detector is encoded in the absorption coefficient $\alpha(k)$ for particles of momentum $k$, and all information about the barrier is contained in the complex amplitude

 \begin{eqnarray}
A_k = \frac{T_k -w_k R_k}{1 - w_k^2}, \label{ak}
\end{eqnarray}

In absence of the potential barrier, $A_k = 1$. Then, Eq. (\ref{plt}) reduces to the time-of-transit distribution for free relativistic particles derived in Ref. \cite{AnSav12}.

\section{Eliminating the superluminality paradox}

We consider the probability distribution $P(L,t)$, Eq. (\ref{plt}) for an initial state $\psi_0$ localized at $x = -x_0$ and with a narrow momentum distribution centered around $k$. In absence of the barrier, $P(L,t)$ is sharply localized around the mean value $(L+ x_0)/v_k$ of the transit time.  The insertion of the barrier induces a delay $t_d(k)$ to the mean transit time, where
			
\begin{eqnarray}
t_d(k) =  \frac{1}{v_k} Im  \frac{\partial  \log A_k}{\partial k}. \label{delt}
\end{eqnarray}

For parity-symmetric potentials, $A_k = T_k$. The time delay of Eq. (\ref{delt}) then reduces to the phase delay time of Bohm and Wigner \cite{bohm, wigner}. Note that Eq. (\ref{delt}) follows from a stationary phase approximation to the probability density $P(L, t)$ of Eq. (\ref{plt}) that is valid only if $P(t)$ has a single maximum.

In general, the temporal properties of the tunneling process are encoded in the full probability distribution $P(L, t)$ associated to a particular experiment, and not in a single parameter like the delay time Eq. (\ref{delt}). The structure of $P(L, t)$ is the only criterion for whether the delay time is meaningfully defined, or for the existence of other, physically significant, timescales. Indeed, if the probability distribution $P(L, t)$ is not characterized by a single peak, the identification of $t_d(k)$ as a delay time is misleading, if not downright wrong \cite{AnSav12b}.

Nonetheless, Eq. (\ref{delt}) applies to several important cases, such as tunneling through a square barrier. In this case, the potential is parity symmetric. This implies that  $A_k = T_k$, and the time delay Eq. (\ref{delt}) reduces to the standard phase delay time \cite{rev}. Our results then still face the challenge of potential superluminal signals. For an opaque square barrier, we recover the standard form for the delay time $t_d(k) = -d/v_k + F(k)$ \cite{rev}, where $F(k)$ is a function only of the incoming particle's momentum $k$. If one defines the time $\tau$ that the particle spent inside the barrier as $\tau = t_d + d/v_k$, then $\tau$ is independent of the barrier length $d$. This implies superluminal traversal velocity for   long barriers.

The problem with this line of reasoning is that the time delay  $t_d(k)$   is not a genuine quantum observable. It does not correspond to a self-adjoint operator (or a positive-operator-valued measure), and it  does not correspond a measurement record in the experiment. The measurement records correspond to the time of transit $t$, not to $t_d$. The time delay is a parameter of the probability distribution for the detection times and it can only be identified after the probability $P(L,t)$ has been determined.
In fact, the identification of the the time delay  $t_d(k)$ requires the combination of data from two different time-of-transit experiments, one with and one without the barrier. Each experiment records a probability distribution for the transit times and $t_d(k)$ is identified as the difference of the corresponding mean values. For this reason, an inference of superluminality from the values of the delay time $t_d(k)$  involves an inadmissible treatment of measurement records that violates the complementarity principle. In no experiment is an actual superluminal signal recorded. The events of particle emission and particle detection are time-like separated. Their proper distance $\Delta s^2 = [(L+x_0)/v_k + t_d]^2 - (L + x_0)^2$ is positive in the regime $L >> d$ where Eq. (\ref{delt}) applies. This is because for $L >> d$ the particle spends considerably more time outside the barrier region.

The positivity of  $\Delta s^2$ for $L >> d$ relies on the fact that for the square barrier,
\begin{eqnarray}
t_d(k) > - d/v_k.  \label{bound}
 \end{eqnarray}
 However, the conclusion is not restricted to this particular case. For $L >> d$, tunneling can be described as a scattering process, and the derivatives of the scattering phase shifts must have a lower bound in order for the $S$-matrix to be compatible with causality \cite{wigner, phshift}. The lower limit in the phase shifts implies a lower bound to the delay time $t_d(k)$. For non-relativistic particles and in absence of bound states,  the lower bound Eq. (\ref{bound}) applies \cite{mugashift}. In general, the corrections to this lower bound are of the order of the particle's de-Broglie wave-length divided by $v_p$, i.e., they are microscopic. They do not affect the argument about the time-like separation between the events of particle emmission and particle detection.

Thus, the only conceivable way of recording a superluminal signal would be in an experimental set-up where both emitter and detector are placed very close to the barrier so that $\Delta s^2 <0$. This implies that $ r/v_p + t_d(p) < r$, where we write $r = L+ x_0$ for the source-detector distance. Since $t_d(p) > -d/ v_p$, the necessary condition for $\Delta s^2 < 0$ is that $d < r < r/(1 - v_p)$. The velocity $v_p$ is bounded above by $\sqrt{1 - (1 + \frac{V_0}{m})^{-2}}$: for larger velocities there is no tunneling.  The approximation of a background electric field fails for $V_0 > m$, because in this case the Hamiltonian Eq. (\ref{hamilton}) would possess negative-energy eigenstates. This implies an upper bound to the velocity  $v_p < \sqrt{3}/2 \simeq 0.87$. A less stringent upper bound to the velocity ($v_p < 2\sqrt{2}/3 \simeq 0.94$) follows from the requirement that the particle's kinetic energy is less than $2m$, so that particle-antiparticle pairs are not created spontaneously. Thus, we conclude that the only conceivable range of values for $r$ where a superluminal might be possible on the basis of Eq. (\ref{delt}) is
 \begin{eqnarray}
 d < r < 7.5 d,
 \end{eqnarray}
 i.e., the   emitter-detector distance $r$ is of the order of the barrier length $d$. However, in this case the  stationary phase approximation that leads to the delay time Eq. (\ref{delt}) is invalid. The position spread $\sigma_x$ of the initial wave-packet must be much smaller than $r$.  To see this, note that the spread $\sigma_x$ in position of the initial wave-packet must be much smaller than $r$. For $r \sim d$, this implies that $\sigma_x >> d$. Hence, the momentum spread $\sigma_p$ must be very large, $\sigma_p >> 1/d$, a condition that is incompatible with the stationary phase approximation employed in the derivation of Eq. (\ref{delt}). The probability distribution $P(L, t)$ is strongly deformed and there is no meaningful definition of a delay time $t_d(k)$.

In fact, the absence of superluminal signals is guaranteed by Eq. (\ref{prob1}) that relates the probability distribution $P(L, t)$, to the correlation functions of a local quantum field. A single-particle  state $| \psi_0 \rangle $ can be expressed as $| \psi_0 \rangle = \int dx \hat{\phi}_1(x) f^*(x)|0\rangle$, for some localized function $f(x)$. For a single-particle initial state localized around $x = -x_0 < -d/2$, Eq. (\ref{prob1}) implies that the probability $P(L, t)$ is proportional to the four-point function
 \begin{eqnarray}
 \langle 0 | \hat{\phi}_1(-x_0,0) \hat{\phi}_1^{\dagger}(L,t) \hat{\phi}_1(L,t) \hat{\phi}_1^{\dagger}(-x_0,0)|0\rangle, \label{4pt}
 \end{eqnarray}
 modulo spatial smearing at the points of emission and detection. Since the points $x = -x_0$ and $x = L$ are outside the barrier region, the field $\hat{\phi_1}$ is effectively free at these points, and they relate to the asymptotic fields defined at $t \rightarrow \pm \infty$ via evolution through the free field Hamiltonian. In particular, $\hat{\phi}_1(-x_0, 0)$ corresponds to the in-field and $\hat{\phi}_2(L, t)$ to the out field of the $S$-matrix formalism. For free fields, the four point function Eq. (\ref{4pt}) factorizes into a product of two point functions. Hence,
 \begin{eqnarray}
 P(L, t) \sim | \Delta(-x_0, L; t)|^2 \label{estim}
 \end{eqnarray}
 where $\Delta(x,x';t) = [\hat{\phi}_1(L,t), \hat{\phi}_1^{\dagger}(-x_0, 0)] = \langle 0|\hat{\phi}_1(L,t) \hat{\phi}_1^{\dagger}(-x_0, 0)|0\rangle$. This means that the probability $P(L,t)$ involves propagation of the initial state through the two-point function $\Delta(x,x';t)$ of the quantum field. In any causal QFT, this Green's function, constructed from the $in$ and $out$ fields {\em must vanish outside the light-cone}, for otherwise it would lead to a non-causal S-matrix.  It follows that the detection probabilities cannot involve superluminal signals. Thus, we claim that the QFT description of tunneling time, enabled through the use of the QTP method, eliminates the superluminality paradox.

The same point of principle holds for the electromagnetic analogues of tunneling, mentioned earlier, where superluminal group velocities have been recorded.  The analogy of those experiments to quantum tunneling is based on the following correspondence. The classical  Helmholtz equation for an electromagnetic field mode $\tilde{E}_{\omega}$ at frequency $\omega$ in an inhomogeneous dielectric medium of refraction index $n(x)$ is
\begin{eqnarray}
\partial_x^2 \tilde{E}_{\omega} + \left(n(x)\omega\right)^2 \tilde{E}_{\omega} = 0.
\end{eqnarray}
This is formally analogous to the time-independent Schr\"odinger equation in presence of a potential
\begin{eqnarray}
\partial_x^2 \psi + \left[2m(E-V(x))\right] = 0,
\end{eqnarray}
if we set $n(x) \omega = \left[2m(E-V(x))\right]^{1/2}$. Depending on the dielectric, it
is possible to have evanescent waves, which decay very much like the quantum mechanical wave functions in tunneling.

However, the analogy between the Helmholz and the Schr\"odinger equations holds only if they are viewed as describing classical waves equations. At the quantum level there is a significant difference. Time evolution according to Schr\"odinger's equation is always unitary. However, the effective Hamiltonian operator for the quantum electromagnetic field in presence of a   dielectric is
\begin{eqnarray}
\hat{H}_{eff} = \sum_a n(\omega_a) \omega_a \hat{a}^{\dagger}_a \hat{a}_a, \label{hem}
\end{eqnarray}
where $\hat{a}, \hat{a}^{\dagger}$ are creation and annihilation operators of the electromagnetic field, and $a$ labels the field modes. Evanescent waves appear for imaginary values of the refraction index $n(\omega_a)$, in which case the operator (\ref{hem}) becomes non-hermitian. The effective quantum evolution according to Eq. (\ref{hem}) would then be non-unitary. This is to be expected, since the microscopic mechanism generating an imaginary refraction index is the absorption of photons by the atoms of the medium. This implies that the electromagnetic analogues of tunneling are in fact analogues of time-of-arrival experiments in open quantum systems, where the effects of dissipation and noise have to be incorporated in the quantum description.

A full treatment of this effect requires the application of the QTP method for the treatment of temporal measurements in open quantum systems, which will be the topic of a different publication.
Here, we point out that a fully quantum treatment of the electromagnetic field in inhomogeneous media requires the complete Quantum Electrodynamics (QED) Hamiltonian for the interaction between the electromagnetic field and the medium. That is, we consider a Hilbert space ${\cal H}_{tot} = {\cal H}_{EM} \otimes {\cal H}_{med} \otimes {\cal H}_{app}$, where ${\cal H}_{EM}$ is associated to the electromagnetic field, ${\cal H}_{med}$ to the particles forming the dielectric medium and ${\cal H}_{app}$ associated to the measurement apparatus. The dynamics on ${\cal H}_{EM} \otimes {\cal H}_{med}$ is governed by the QED Hamiltonian
\begin{eqnarray}
\hat{H}_{QED} = \hat{H}_{EM} \otimes 1 + 1 \otimes \hat{H}_{med} + \int dx \hat{A}^i(x)\otimes \hat{j}_i(x),
\end{eqnarray}
where $\hat{H}_{EM} $ is the Hamiltonian for the free electromagnetic field, $\hat{H}_{med}$ is the self-Hamiltonian for the particles in the medium, $\hat{A}^i(x)$ the electromagnetic potential and $\hat{j}_i$ the current associated to the medium.

The coupling of the electromagnetic field with the detector is governed by an interaction Hamiltonian $\hat{H}_I$ analogous to Eq. (\ref{prob1})
\begin{eqnarray}
\hat{H}_I = \int dx \hat{A}^i(x) \hat{J}_i(x),
\end{eqnarray}
where $\hat{J}_i(x)$ is the electric current associated to the detector degrees of freedom. We can then derive an equation for the probability density of detection $P(L,t)$ analogous to Eq. (\ref{prob1})
\begin{eqnarray}
P(L, t) = \int d\tau \int dz R^{ij}(z, \tau)   \langle \psi_0, \Omega|\hat{A}_i^{\dagger}(L -\frac{z}{2}, t-\frac{\tau}{2}) \hat{A}_j(L+\frac{z}{2}, t+\frac{\tau}{2})|\psi_0, \Omega\rangle, \label{prob1f}
\end{eqnarray}
where $R^{ij}(\tau, z)$ is a kernel incorporating the detector degrees of freedom, and $|\psi_0, \Omega \rangle = |\psi_0\rangle \otimes |\Omega\rangle$, where $|\psi_0 \rangle$ is the initially prepared electromagnetic field state, and $|\Omega$ the initial state of the particles in the medium (for example, an energy eigenstate).
Using the same arguments leading to the estimation Eq. (\ref{estim}) for $P(L,t)$, we can show that the propagation of the electromagnetic field signal in Eq. (\ref{prob1f}) is guided by the two-point function of the field $\hat{A}_i$ that corresponds to photon emission and detection. If QED is a consistent quantum field theory these functions must be causal, irrespective of   the initial state $|\Omega\rangle $ of the particles in the dielectric medium.
 Thus, no superluminal signal is to be expected in transit-time measurements of photons through an absorbing medium.

\section{Conclusions}

To summarize, our proposed resolution to the superluminality paradox is the following. The time delay $t_d(k)$ can indeed be determined in {\em some} time-of-transit experiments on tunneling set-ups. However, it is not a genuine quantum observable, but a parameter of the probability distribution $P(L,t)$. An inference of superluminal velocities in the tunneling region (or indeed any velocity) from the value of $t_d(k)$ involves classical reasoning that is incompatible with the rules of quantum theory. All physical information about signal propagation is contained in the  probability distribution $P(L,t)$, Eq. (\ref{prob1}), which is guaranteed to be causal, because, the QTP method enables its definition  in terms of the correlation functions of a local quantum field theory.

In this sense, the apparent superluminality paradox in tunneling is a non-trivial manifestation of the particle-wave duality in quantum theory. Group velocities may be greater than the speed of light, but they are defined in terms of set-ups that measure wave properties of the quantum field, and not transit times. The latter correspond to the particle aspects of the quantum field, and they are measured in different experiments. Inferences of superluminality follow from attempts to relate observables defined in different experiments.   Indeed, the only way to relate physical magnitudes associated to different experiments is through the introduction of hidden variables, i.e., variables describing microscopic properties of the system that are not contained in the formalism of standard quantum theory. But superluminality would hardly be surprising in this case. Superluminal velocities are to be expected in any hidden-variables theory that reproduces the predictions of quantum mechanics, by virtue of Bell's theorem \cite{Bell}.

 For this reason, we believe that time-of-transit experiments in tunneling systems could be of great significance for the foundations of quantum mechanics. We expect that the probability distribution Eq. (\ref{plt}) for the time of transit is incompatible with any local hidden-variable theory, i.e., a hidden-variable theory that admits no superluminal propagation. Thus, experiments aiming to the confirmation of Eq. (\ref{plt}) could provide a new ground for testing the predictions of quantum theory Vs. local realism. They could demonstrate that quantum `non-locality' refers not only to the correlations between quantum subsystems, but also to the values of temporal observables in non-composite quantum systems.

\end{document}